# Thickness-Driven Quantum Anomalous Hall Phase Transition in Magnetic Topological Insulator Thin Films

Yuchen Ji,[1,2†] Zheng Liu,[3†] Peng Zhang,[4†] Lun Li,[2,5] Shifei Qi,[3,6,] Peng Chen,[2,5] Yong Zhang,[2,5] Qi Yao,[1] Zhongkai Liu,[1] Kang L. Wang,[4] Zhenhua Qiao,[3*] and Xufeng Kou[1,5*]

[1]ShanghaiTech Laboratory for Topological Physics, School of Physical Science and Technology, ShanghaiTech University, Shanghai, China, 200031

[2]University of Chinese Academy of Sciences, Beijing, China, 101408

[3]International Center for Quantum Design of Functional Materials, Hefei National Laboratory for Physical Sciences at Microscale, CAS Key Laboratory of Strongly-Coupled Quantum Matter Physics, and Department of Physics, University of Science and Technology of China, Hefei, Anhui, China, 230026

[4]Department of Electrical Engineering, University of California, Los Angeles, California, USA, 90095

[5]School of Information Science and Technology, ShanghaiTech University, Shanghai, China, 20031

[6]College of Physics, Hebei Normal University, Shijiazhuang, Hebei, China, 050024

†These authors contributed equally to this work.

*To whom correspondence should be addressed. qiao@ustc.edu.cn; kouxf@shanghaitech.edu.cn

The quantized version of anomalous Hall effect realized in magnetic topological insulators (MTIs) has great potential for the development of topological quantum physics and low-power electronic/spintronic applications. To enable dissipationless chiral edge conduction at zero magnetic field, effective exchange field arisen from the aligned magnetic dopants needs to be large enough to yield specific spin sub-band configurations. Here we report the thickness-tailored quantum anomalous Hall (QAH) effect in Cr-doped $(Bi,Sb)_2Te_3$ thin films by tuning the system across the two-dimensional (2D) limit. In addition to the Chern number-related metal-to-insulator QAH phase transition, we also demonstrate that the induced hybridization gap plays an indispensable role in determining the ground magnetic state of the MTIs, namely the spontaneous magnetization owing to considerable Van Vleck spin susceptibility guarantees the zero-field QAH state with unitary scaling law in thick samples, while the quantization of the Hall conductance can only be achieved with the assistance of external magnetic fields in ultra-thin films. The modulation of topology and magnetism through structural engineering may provide a useful guidance for the pursuit of QAH-based new phase diagrams and functionalities.

When a perpendicular magnetic moment is introduced into the host topological insulators, the broken time-reversal symmetry (TRS) would lift the spin degeneracy and give rise to the formation of massive Dirac fermions of the surface states[1-5]. Once the spin splitting is finely adjusted with respect to the intrinsic spin-orbit coupling strength, the constituent effective exchange field would cause band inversion in one set of the spin sub-bands while maintaining the other one in the topologically trivial regime (Fig. 1a)[2, 3]. As a result, this crucial band structure leads to the formation of the QAH effect (non-zero Chern number) in MTIs, where scale-invariant chiral edge-state current can conduct coherently without energy dissipation at zero magnetic field[6,7]. Such salient quantum transport features, along with versatile magnetic order-driven topology manipulation strategies, have resulted in unprecedented advancements of TRS-breaking topological quantum physics in the past decade[8-17].

Experimentally, the adoption of molecular beam epitaxy (MBE) has enabled the growth of high-quality Cr/V-doped $(Bi,Sb)_2Te_3$ thin films to explore QAH-related electronic transport phenomena[18-22]. In these MTI systems, the magnetic dopants provide robust long-range ferromagnetic order, and appropriate Bi-to-Sb ratio ensures the Fermi level position within the energy gap of the surface states. Moreover, when the film thickness is curtailed into the 2D region, the hybridization between the top and bottom surfaces induces a topologically trivial gap that may change the Chern number ($C$), and such $C$-parity change serves as the prerequisite for both the realization of metal-to-insulator QAH phase switching and the generation of quasi-particles such as axions and Majorana fermions[23-26]. Accordingly, the key to implement the aforementioned exotic phenomena lies in the MTI structural engineering. On one hand, the dimensional reduction can enlarge the hybridization gap (i.e., the $C = 0$ phase); on the other hand, the magnetic exchange coupling and microscopic spin textures are also found to be sensitive to film thickness variation[27-29]. For instance, it is suggested that the presence of non-mean-field-like ferromagnetism and increased magnetic disorders may cause the transport in ultra-thin samples to deviate from the ideal quantization state. In the extreme case where the MTI film thickness is below 4~5 quintuple-layers (QLs), a phenomenological picture from experiments indicates that the band inversion condition may no longer be satisfied, hence restricting the system within the insulating phase at deep cryogenic temperatures.

To elucidate the above proposal, in this *Letter*, we prepare a set of Cr-doped $(Bi,Sb)_2Te_3$

thin films across the 3D-to-2D regions, with which critical responses of the magnetic exchange energy and hybridization gap to the film thickness are quantitatively investigated. Depending on the initial band topology, the QAH state evolved from the non-trivial topological phase (i.e., both the top and bottom surface states are well-defined) warrants macroscopic dissipationless chiral edge conduction at zero field, whereas large external magnetic field is required to convert the 2D normal insulators (i.e., ultra-thin MTI samples) into the $C = \pm 1$ states with resumed fully-quantized Hall plateau. Furthermore, the visualized thickness-sensitive QAH phase diagrams can be well-explained by the explicit first-principles calculation, which unveils diminished Van Vleck spin susceptibility and weakened ferromagnetism with the decrease of the film thickness. Our results not only identify the magnetic origins of the MTI samples, but also manifest the importance of tuning parameters on constructing robust QAH states and topological order transition-related phenomena.

**Results**

**Thickness-tailored quantum anomalous Hall phase diagram**

High-quality $(Cr_{0.2}Bi_{0.12}Sb_{0.68})_2Te_3$ samples with film thickness ($d$) ranging from 4 to 7 QLs were epitaxially grown on the semi-insulating GaAs(111)B substrates based on the optimized MBE growth procedure established in our previous work[19,24,28]. The film stoichiometry (i.e., Bi-to-Sb ratio) was carefully calibrated so that the Fermi level ($E_F$) was positioned inside the surface gap without additional gate tuning (Fig. S1), as confirmed by the angle resolved photoemission spectroscopy (ARPES) in Fig. 1b. In the meantime, *in-situ* reflection high energy electron diffraction (RHEED) technique was used for the precise control of the film thickness (Fig. 1c). Consequently, these samples provide a reliable material platform to explore the dimension-correlated QAH scaling behaviors.

After sample growth, standard four-point magneto-transport measurements were performed on the $(Cr_{0.2}Bi_{0.18}Sb_{0.62})_2Te_3$-based mm-size Hall bar devices with $d$ = 4, 5, 6, 7 QLs, respectively. As illustrated in Figs. 2a-2d, the corresponding magnetic field dependence of longitudinal resistance ($R_{xx}$) and Hall resistance ($R_{xy}$) at $T$ = 50 mK can be divided into two distinct categories regarding the film thickness. Specifically, both the square-shape hysteresis Hall loops with quantized value of $R_{xy} = \pm h/e^2$ (where $h$ is the Planck constant, $e$ is the electron charge, and the sign is decided by the chirality of the edge conduction with respect to the magnetization direction) and vanishing $R_{xx}$ at zero magnetic field are observed in $d$ = 6 and 7 QLs samples, demonstrating the realization of QAH effect. Conversely, the anomalous Hall resistance in the $d$ = 4 or 5 QLs films exhibits a non-saturation feature in the low field regime (Fig. S2), and the quantization of $R_{xy}$ is not achieved until $B > 10$ T. Here, we emphasize that such high-field-enabled quantization revealed in our ultra-thin MTI thin film samples has a different physical origin from the quantum Hall effect because the Fermi level is constantly located inside the band gap without being moved across the discrete Landau levels[30, 31]. Instead, the occurrence of the QAH chiral edge state assisted by the high magnetic field may suggest an intricate interplay between the magnetism and the band structure, as will be elaborated later.

Strikingly, the 3D-to-2D QAH evolution becomes more evident when the magneto-transport data are presented in the conductance plots (Figs. 2e-2h). Consistent with previous reports[24, 25], when the hybridization gap is introduced in the 6 QLs MTI thin film samples, it allows for the metal-to-insulator QAH phase switching, where two zero Hall plateau are developed around the coercivity fields and the longitudinal conductance $\sigma_{xx}$ goes through two peaks separating the $C = \pm 1$ and 0 states (Fig. 2g). As the film thickness further decreases, the dilated hybridization between the top and bottom surfaces, along with suppressed bulk conduction, is found to be responsible for the formation of the insulating state with higher $R_{xx}$ peak amplitude and broadened intermediate Hall mesa. In the low-thickness limit case ($d$ = 4 QLs), the $C$ = 0 phase becomes dominant over a wide magnetic field range (i.e., [−5, 5 T]), and the transition of ($\sigma_{xx}$, $\sigma_{xy}$) from (0, 0) to (0, ±$e^2$/$h$) endpoints discloses a smoother trend as compared with thicker MTI thin films (Fig. S3). These findings thus highlight the effectiveness of the accurate thickness control on the manipulation of the QAH phases with different topological orders.

In light of the importance of the dimensional effect on the QAH phase transition, we subsequently performed the field-cooling measurements on the MTI films. Governed by the chiral edge conduction nature of QAH, the longitudinal resistance $R_{xx}$ is expected to decrease as the system approaches the QAH state at lower temperature[32]. In the case of $d$ ≥ 6 QLs (Cr$_{0.2}$Bi$_{0.12}$Sb$_{0.68}$)$_2$Te$_3$ samples, the edge channels with $C$ = ±1 are formed by the ferromagnetic order below Curie temperature. Consequently, the temperature-dependent $R_{xx}$ data display almost identical scaling slopes, regardless of the applied magnetic field (Fig. 3c). On the contrary, the $R_{xx}(B)$ – $T$ curves of the 4 QL sample show a strong field-dependent metallic-to-insulating phase crossover. In particular, as the sample gradually cools down to deep cryogenic temperatures, $R_{xx}$ increases monotonically in the low field regime (i.e., signifying the diffusive transport characteristic of a trivial insulator), while its magnitude drops towards zero under the high-field cooling (i.e., in reference to the QAH conduction scenario); at the boundary of $B$ = 5.5 T, the measured $R_{xx}$ remains almost constant in the entire 50 mK< $T$ < 2 K range (Fig. S4a). Likewise, a similar $R_{xx}(B)$ – $T$ contour, with a smaller critical transition field of $B$ = 0.15 T, is also observed in the 5 QLs sample, as shown in Fig. 3b.

The decisive influence of dimensional reduction on tailoring the QAH phase is further visualized by the renormalization group (RG) flow diagrams[33, 34]. Figures 3d-3f summarize the positive-field-cooled ($\sigma_{xy}(T)$, $\sigma_{xx}(T)$) plots for the same three MTI samples examined above. As the temperature decreases, all ($\sigma_{xy}(T)$, $\sigma_{xx}(T)$) trajectories of the $d = 6$ QL sample tend to converge to the fixed QAH point at ($e^2/h$, 0), whereas the developed conductance data points of the $d = 4$ and 5 QLs films flow to cover the continuous (0, 0)-to-($e^2/h$, 0) semicircle, again manifesting the magnetic field-assisted topological phase transition under the framework of the global QAH phase diagram (i.e., the slight deviation from the ideal semi-circular curve is due to the finite $R_{xx}$ detected at 50 mK). Interestingly, the magnetic field-trained RG flows in Figs. 3d-3e resolve an unusual asymmetric pattern with the unstable fixed point at ($0.26e^2/h$, $0.44e^2/h$) instead of the ($0.5e^2/h$, $0.5e^2/h$) value predicted by the modular symmetry group theory[35]. It is noted that similar ($B$-fixed, $T$-dependent) RG flow asymmetry has been observed in strongly-disordered graphene at low temperatures[36]; meanwhile the position of the unstable point is affected by the spin-splitting situation at the transition boundary[35]. Both arguments may also be applied to our MTI thin film samples with low carrier mobility, highly-insulating resistance, and attenuated magnetic ordering strength. Nevertheless, the exact mechanism and related model need to be further justified.

**Thickness-dependent magnetism by first-principles calculation**

To understand the thickness-driven QAH phase transition discovered in Figs. 2 and 3, we first recall that when the MTI samples are below 6 QLs, the hybridization between the top and bottom surfaces drives the ultra-thin films into the topologically trivial sate, and a strong magnetic exchange coupling is necessary to overcome the resulting hybridization gap so that one spin sub-band can be inverted[2,3]. To quantitatively correlate such two-dimensional quantum confinement with our experimental observations, we calculated the critical exchange field ($\Delta_C$) that enables the inversion of the $(Bi,Sb)_2Te_3$ band structure by using the tight-binding model (Fig. S5)[37,38]. As displayed in Fig. 4b, it shows that $\Delta_C$ experiences a sharp increase from 0.07 eV to 0.16 eV as the film thickness reduces from 6 QLs to 4 QLs; in other words, the QAH chiral edge conduction becomes much more difficult to achieve in thinner MTI films. In accordance with the QAH phase diagram with universal scaling law in Figs. 3c and 3f, our numerically obtained critical exchange field also predicts that the topologically trivial gap will be closed when $d > 6$ QLs and the thick MTI samples always preserve the $C = \pm 1$ signature.

More importantly, the decrease of the film thickness not only enlarges the hybridization gap, but also weakens the intrinsic magnetic exchange coupling of the MTI system. In general, when the samples reach the QAH state at deep cryogenic temperatures, the absence of bulk conduction means that the Cr $d$-orbit moments can only be locally aligned by the spin susceptibility ($\chi_e$) of the band electrons via the Van Vleck mechanism whose expression is described as[3]

$$\chi_e = \frac{1}{N}\sum_k \chi_k, \ \ \chi_k = 4\mu_0 (\frac{\mu_B}{\hbar})^2 \sum_{E_{nk} < E_F < E_{mk}} \frac{\langle nk|S_z|mk\rangle\langle mk|S_z|nk\rangle}{E_{mk} - E_{nk}}$$

where $\mu_0$ is the vacuum permeability, $\mu_B$ is the Bohr magneton, $\boldsymbol{S_z}$ is the spin operator, $|mk\rangle/E_{mk}$ and $|nk\rangle/E_{nk}$ are the Bloch functions/eigenstate energies of the conduction and valence bands, respectively. The summation includes uniformly distributed $k$ points in the first Brillouin zone, and $N$ represents the number of $k$ points. Accordingly, it is proposed that the mixing between the inverted Bi/Sb $p_{1z}$-state ($|nk\rangle$) and Te $p_{2z}$-state ($|mk\rangle$) determines the overall coupling strength ($\langle nk|\boldsymbol{S_z}|mk\rangle$). Following such scenario, thickness-dependent $\chi_e$ of the particular (Bi,Sb)$_2$Te$_3$ system was numerically estimated (Fig. S6), and the obtained positive correlation of the $\chi_e - d$ curve (blue squares in Fig. 4c) confirms the essential impact of the non-trivial band structure on the magnetic property of the MTI system. Additionally, the projected augmented-wave method implemented in the Vienna *ab initio* simulation package (VASP) was further employed to

numerically investigate the magnetic order of the exact $[2 \times 2 \times n]$ Cr-doped $(Bi,Sb)_2Te_3$ supercell (i.e., where n is the number of QLs)[39, 40]. Similar to Van Vleck spin susceptibility, the energy difference between the antiferromagnetic (AFM) and ferromagnetic (FM) configurations of the doped Cr elements in the $(Bi,Sb)_2Te_3$ matrix presents a monotonic downward trend with the decrease of the film thickness (red circles in Fig. 4c), and its corresponding strength is significantly smaller than the band inversion-needed value of $\Delta_C$ in the $d < 6$ QLs region. Considering that the overall exchange field of the MTI system is expressed as $\Delta = \Delta_0 + \eta \chi_e B$ (where $\Delta_0 = 98$ meV represents the spontaneous exchange field introduced by the Cr dopants at zero magnetic field, and $\eta$ is the coefficient that can be fitted from the our experimental data), we therefore are able to establish the relationship between the exchange field and the applied external magnetic field. As highlighted in Fig. 4d, the critical magnetic field that triggers the QAH phase transition decreases remarkably as the MTI thickness increases from 4 QLs to 6 QLs, and the consistency between the calculated onset magnetic field (red spheres) and the thickness-tailored QAH phase diagram hence uncovers the underlying physics of the field-induced quantization in the ultra-thin MTI samples.

**Discussions**

On the basis of the experimental results and theoretical calculations, we may conclude that for thick MTI samples ($d \geq 6$ QLs), both the large electron spin susceptibility and the preferable parallel configuration of Cr atoms lead to a stable ferromagnetic ground state that automatically maintains the spin sub-band inversion condition without the help of magnetic field. In contrast, the appearance of sizable hybridization gap would drive the ultra-thin films ($d = 4$ and 5 QLs) into the trivial insulator phase in the low-field region. Under such circumstances, the subtle $\chi_e$ strength due to reduced band mixing cannot sustain the ferromagnetic coupling among Cr dopants, and a strong external magnetic field is therefore required to generate sufficient exchange field ($\chi_e B$) so as to facilitate the completion of the QAH insulator-to-metal phase transition. Besides, we need to point out that such thickness-modulated magnetism is only observed in the QAH state; when the base temperature is elevated to 1.5 K (i.e., diffusive transport regime), hole-mediated Ruderman−Kittel−Kasuya−Yosida (RKKY) mechanism is expected to dominate the magnetic exchange interaction[41, 42]. As a result, all the $p$-type Cr-doped (Bi,Sb)$_2$Te$_3$ samples, with identical magnetic doping level and carrier density, exhibit a thickness-independent anomalous Hall effect featured by the same coercive and saturating magnetic fields, as shown in Fig. S7.

In conclusion, we study the QAH phase transition in the MBE-grown (Cr$_{0.2}$Bi$_{0.12}$Sb$_{0.68}$)$_2$Te$_3$ thin films. It shows that the change of film thickness would affect both the topological and the magnetic orders of the as-grown samples. Given that the hybridization gap would transform ultra-thin MTI film into a trivial insulator with negligible remnant magnetic moment, the zero-field quantization cannot be realized, and such criteria set the lower thickness limit of QAH in the Cr-doped (Bi,Sb)$_2$Te$_3$ system. On the other hand, the capability to modify the zero Hall plateau width (i.e., $C = -1 \leftrightarrow 0 \leftrightarrow +1$ transition) by varying the film thickness in the 2D regime may provide an effective approach to explore the chiral Majorana edge mode and axion insulators, hence enriching the opportunities of MTI-based applications.

## Methods

**MBE growth.** MTI thin film growth was carried out on the epi-ready semi-insulating GaAs(111)B substrates by MBE (base pressure below $1 \times 10^{-10}$ mbar). Prior to sample growth, the GaAs substrate was pre-annealed at 570°C under the Se-protected environment in order to remove the native oxide surface. During the (Cr$_{0.2}$Bi$_{0.12}$Sb$_{0.68}$)$_2$Te$_3$ thin film growth, high-purity Cr (99.99%) and Bi (99.9999%) atoms were co-evaporated from standard Knudsen cells, while Sb (99.99999%) and Te (99.99999%) were evaporated by thermal cracker cells. The element ratio was calibrated by the beam-flux monitor before sample growth, and epitaxial growth was monitored by an *in-situ* RHEED technique by which digital RHEED images were captured using a KSA400 system built by K-space Associates, Inc. After sample growth, a thin Te capping layer was deposited to protect the surface states.

**ARPES**. After the as-grown $(Cr_{0.2}Bi_{0.18}Sb_{0.62})_2Te_3$ sample being loaded into the ARPES system, the Te-capping layer was removed in the preparation chamber until sharp streaky 2D RHEED pattern resumed. The ARPES measurements were subsequently carried out at 80K by using a Scienta R4000 electron energy analyzer. A Helium discharge lamp with a photon energy of $hv = 21.218$ eV was used as the photon source, and the energy resolution of the electron energy analyzer was set at 15 meV.

**Magneto-transport measurements**. The MTI thin films were manually etched into a six-probe Hall bar geometry with typical dimensions of $2 \times 1$ mm$^2$. The electrodes were made by welding indium shots onto the contact areas of the thin film. The magneto-transport measurements were performed with a He-4 refrigerator (Oxford TeslatronPT system). Several experimental variables such as temperature, magnetic field, and lock-in frequency were varied during the measurements. Multiple lock-in amplifiers and Keithley source meters (with an AC excitation current of $I = 10$ nA) were connected to the samples to enable the precise four-point lock-in transport experiments.

**First-principles calculation.** First-principles calculations were performed by using the projected augmented-wave method implemented in the Vienna ab initio simulation package (VASP). The generalized gradient approximation (GGA) of Perdew-Burke-Ernzerh of type was used to treat the exchange-correlation interaction. For magnetic property calculation, K-mesh points of $3\times3\times1$ and 5×5×1 were used for the structural relaxation and total energy estimation, respectively. A vacuum space of 20 Å was used to avoid spurious interactions. The kinetic energy cutoff was set to be 400 eV for all the thin-film calculations. All atoms were allowed to relax until the Hellmann-Feynman force on each atom was less than 0.01 eV/ Å. The GGA+U method with U=3.0 eV and J=0.87 eV was used to treat the Coulomb interaction of the Cr element.

In the magnetic property calculation, a 2×2×n $(Bi,Sb)_2Te_3$ supercell was used with n representing the number of QLs. The concentration of Bi element was set to be 12.5%, and two Cr atoms were used to substitute the Sb site of $(Bi,Sb)_2Te_3$ (i.e., same as the as-grown $(Cr_{0.2}Bi_{0.18}Sb_{0.62})_2Te_3$ samples). The magnetic coupling strength of $(Bi,Sb)_2Te_3$ thin films was calculated by comparing the energy difference between antiferromagnetic and ferromagnetic configurations.

**Tight-binding model.** The thickness-dependent critical magnetic exchange field was evaluated by using the tight-binding model Hamiltonian. In the tight-binding representation, the topological insulator doped with magnetic elements can be written as:

$$H = H_{3D} + H_{imp},$$

$$H_{3D} = \sum_i E_0 \phi_i^\dagger \phi_i + \sum_i \sum_{\alpha=x,y,z} \phi_i^\dagger T_\alpha \phi_{i+\hat{\alpha}} + \text{H. c.,} \qquad (1)$$

$$H_{imp} = \sum_i m_0 \phi_i^\dagger \phi_i,$$

where $H_{3D}$ describes the bulk Hamiltonian of the 3D topological insulator, $E_0$, $T_\alpha$, $m_0$ are respectively written as:

$$E_0 = \left(M - \sum_\alpha B_\alpha\right)\sigma_0 \otimes \tau_z - \sum_\alpha D_\alpha \sigma_0 \otimes \tau_0,$$

$$T_\alpha = \frac{B_\alpha}{2}\sigma_0\otimes\tau_z + \frac{D_\alpha}{2}\sigma_0\otimes\tau_0 - \frac{iA_\alpha}{2}\sigma_\alpha\otimes\tau_x, \tag{2}$$

$$m_0 = m\sigma_z\otimes\tau_0,$$

where $M$ represents the inverted band gap and $A_\alpha$ reflects the Fermi velocity, $\hat{\alpha}$ is the unit vector along $\alpha = (x, y, z)$ direction, $\sigma$ and $\tau$ are spin and orbital Pauli matrices, respectively. $H_{\text{imp}}$ is used to describe the magnetic dopants with $m$ representing the effective exchange field strength. The effective Hamiltonian is written in cubic lattice with the electronic state at each site i expressed as $\phi_i = (a_{i\uparrow}, b_{i\uparrow}, a_{i\downarrow}, b_{i\downarrow})$, where $(a, b)$ denote two independent orbitals and $(\uparrow, \downarrow)$ represent spin indices. Since we only concerned about the thickness-dependent topological property of topological thin films, we set the $x$, $y$ direction as periodic directions and z direction are set to be finite. Without loss of generality, we set $A_\alpha = A = 1.5, B_\alpha = B = 1.0, D_\alpha = D = 0.1$, and $M = 0.3$.

**Van Vleck spin susceptibility calculation.** Van Vleck spin susceptibility was calculated by using the following equations:

$$\chi_e = \frac{1}{N}\sum_k \chi_k, \ \ \chi_k = 4\mu_0(\frac{\mu_B}{\hbar})^2 \sum_{E_{nk}<E_F<E_{mk}} \frac{\langle nk|S_z|mk\rangle\langle mk|S_z|nk\rangle}{E_{mk}-E_{nk}} \tag{3}$$

where $\mu_0$ is the vacuum permeability, $\mu_B$ is the Bohr magneton, $S_z$ is the spin operator. $n$, $|nk\rangle, E_{nk}$ represents band index, wavefunction, and eigenvalue of $n$-th band at momentum $k$, respectively. The summation includes uniformly distributed k points in the first Brillouin zone, and $N$ represents the number of $k$ points.

**Acknowledgements:**

We thank Dr. L. Pan, Prof. G. Li, and Prof. Y. G. Yao for helpful discussions. We are grateful for the support from National Key R&D Program of China under contract number 2017YFA0305400, the National Natural Science Foundation of China (NSFC, Grant No. 61874172), and the Strategic Priority Research Program of Chinese Academy of Sciences (Grant No. XDA18010000). Work at USTC was supported by NSFC (Grant Nos. 11974327, 11674024, and 11974098), the Fundamental Research Funds for the Central Universities, and the Anhui Initiative in Quantum Information Technologies. We are grateful to the AMHPC and Supercomputing Center of USTC for providing high-performance computing assistance.

**Author Contributions**

X. K. and Z. Q. conceived and designed the research. Y. J., P. C., and Q. Y. grew the material. Y. J., P. Z., L. L., and Y. Z. performed the measurements. Z. L. and K. L. W. contributed to the measurements and analysis. Z. L., S. Q., and Z. Q. designed the theoretical model. Y. J., Z. Q., and X. K. wrote the paper with help from all of the other co-authors.

**Additional information**

**Supplementary Information** accompanies this paper at

**Competing financial interests**: The authors declare no competing financial interests.

**Reprints and permission** information is available online at

**Figure Captions:**

**Figure 1 | Band structure and characterizations of the MBE-grown $(Cr_{0.2}Bi_{0.12}Sb_{0.68})_2Te_3$ film.**
**a.** Schematic of band structure evolution of TIs with the presence of magnetic exchange coupling. The red and blue cones donate the two sub-bands with opposite spin polarizations. With appropriate exchange field, the QAH state can be either constructed from a normal insulator (NI, left panel) or transformed from the original band inversion state (TI, right panel). **b.** Surface band structure measured by ARPES along the $K$-$\Gamma$-$K$ direction. The blue dashed line indicates the Fermi level position, and the red dashed lines confirm that the Dirac surface states intersect at the Dirac point (DP). **c.** RHEED oscillation showing the precise control of the film thickness ranging from 4 QLs to 7 QLs during thin film growth.

**Figure 2 | Thickness-dependent QAH effect in Cr-doped $(Bi,Sb)_2Te_3$ thin films. a-d.** Magnetic field dependence of longitudinal resistance $R_{xx}$ (red) and Hall resistance $R_{xy}$ (blue) of the samples with film thickness of $d$ = 4, 5, 6 and 7 QLs, respectively. **e-h.** Corresponding magnetic field-dependent conductance $\sigma_{xx}$ (red) and $\sigma_{xy}$ (blue) plots of the four MTI films. At the base temperature of 50 mK, thick samples ($d$ = 6 and 7 QLs) develop robust quantized Hall plateau at zero magnetic field whereas full quantization of $R_{xy}$ and $\sigma_{xy}$ can only be obtained in the high-field region of the thin films ($d$ = 4 and 5 QLs).

**Figure 3 | Temperature/field-driven QAH phase diagram and renormalization group flow of the Cr-doped $(Bi,Sb)_2Te_3$ thin films. a-c.** Magnetic field-dependent mappings of the $R_{xx} - T$ curves of the MTI samples with film thickness of $d$ = 4, 5, and 6 QLs, respectively. Same color code is used for a direct comparison and the black dashed lines indicate the boundary between the QAH and trivial insulator states. **d-f.** Parametric plots of $(\sigma_{xy}(T), \sigma_{xx}(T))$ for the $d$ = 4, 5 and 6

QLs thin films. All data points are collected during the field cooling process. As the temperature gradually decreases from 1.8 K to 50 mK, the RG flow of the $d = 6$ QLs sample tend to converge to the fixed QAH point at ($e^2/h$, 0), while the developed conductance results of the $d = 4$ and 5 QLs films cover the continuous (0, 0)-to-($e^2/h$, 0) semicircle. The black solid semicircles of radius $e^2/h$ centered at (0.5, 0) represent the trajectory of the ideal metal-to-insulator phase transition.

**Figure 4 | Thickness-modulated magnetic parameters of the Cr-doped (Bi,Sb)$_2$Te$_3$ system. a.** Illustration of the [2 × 2 × n] supercell of the Cr-doped (Bi,Sb)$_2$Te$_3$ structure used in the first-principles calculation. The concentration of Bi element is set to be 12.5%, and two Cr atoms are used to substitute the Sb site of the (Bi,Sb)$_2$Te$_3$ matrix in accordance with the MBE-grown samples investigated in this work. **b.** The critical exchange field $\Delta_C$ for realizing the QAH effect as a function of the film thickness based on the tight-binding model estimation. **c.** Both the Van Vleck spin susceptibility (blue squares) and the energy difference between the AFM and FM configurations (red circles) are found to be suppressed as the film thickness decreases from 6 QLs to 4 QLs. The data are computed by using VASP and wannier-90 packages. **d.** Summary of the thickness and field-tailored QAH phase diagram in terms of the measured $R_{xx}$ data points of Fig. 2. The critical magnetic field (red spheres) estimated from the numerical exchange field calculation agrees with the experimental results.

## FIGURE LEGEND

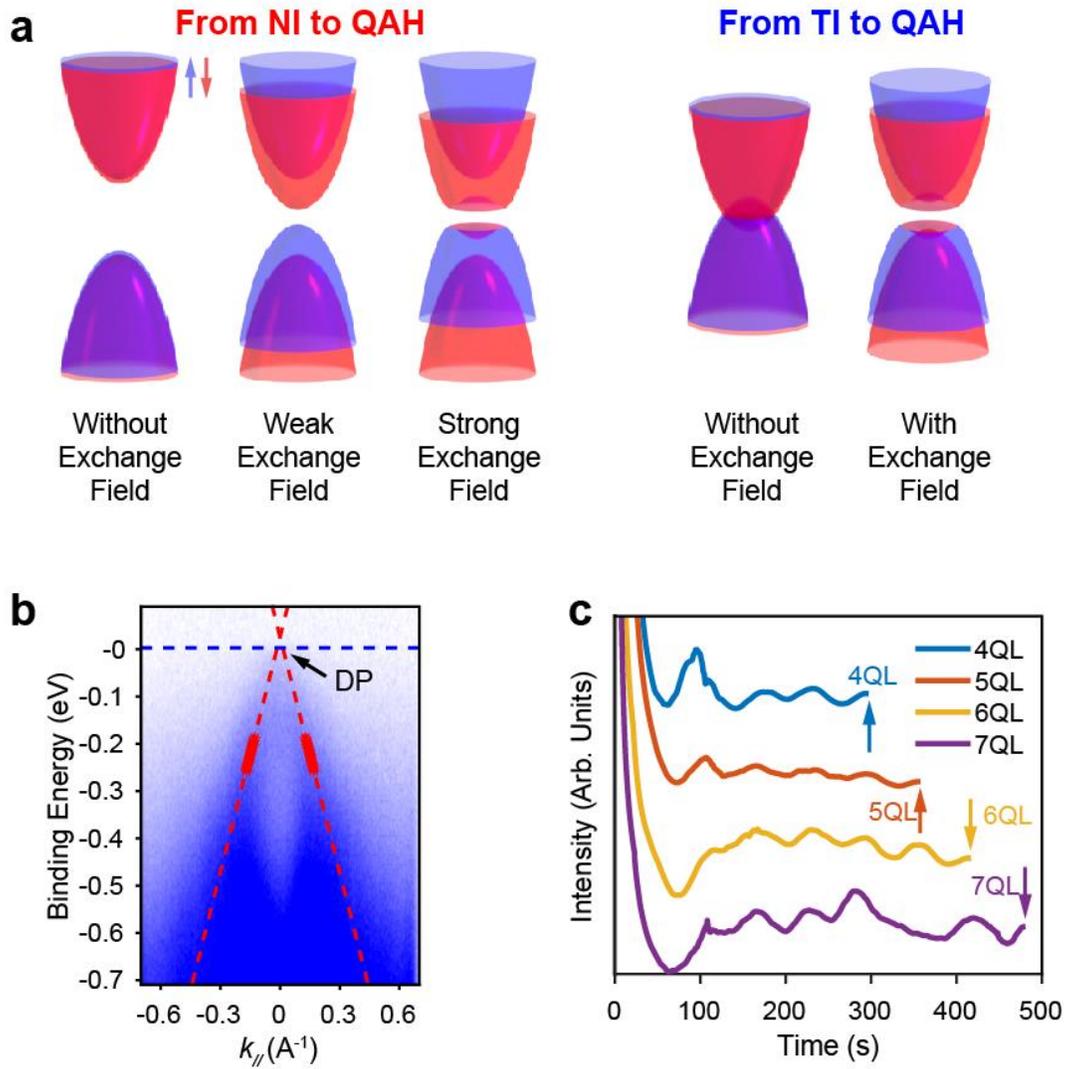

**Figure 1**

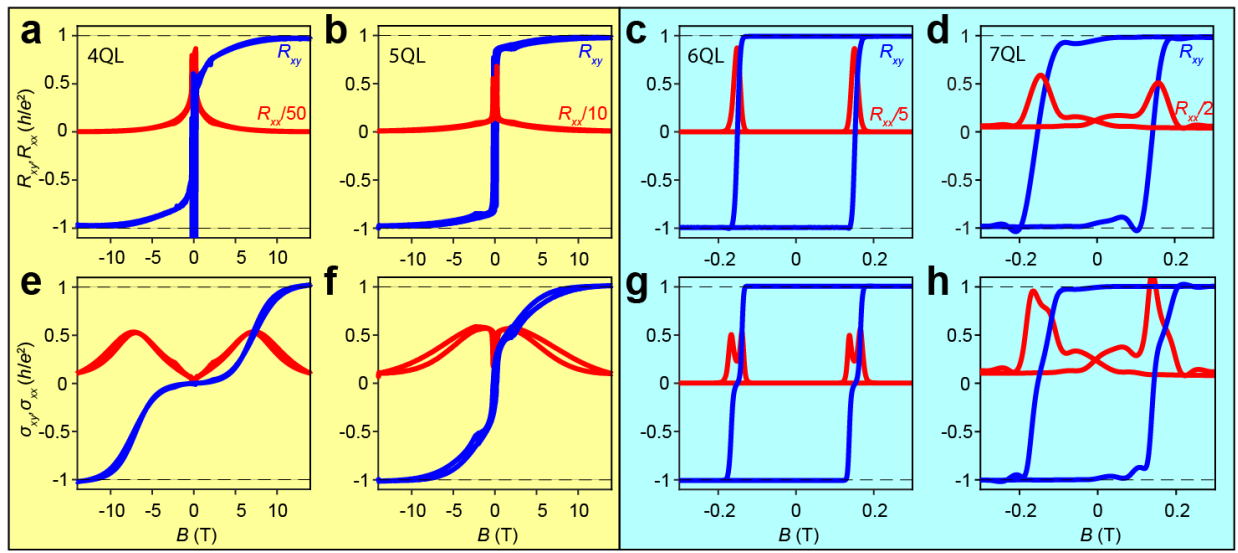

**Figure 2**

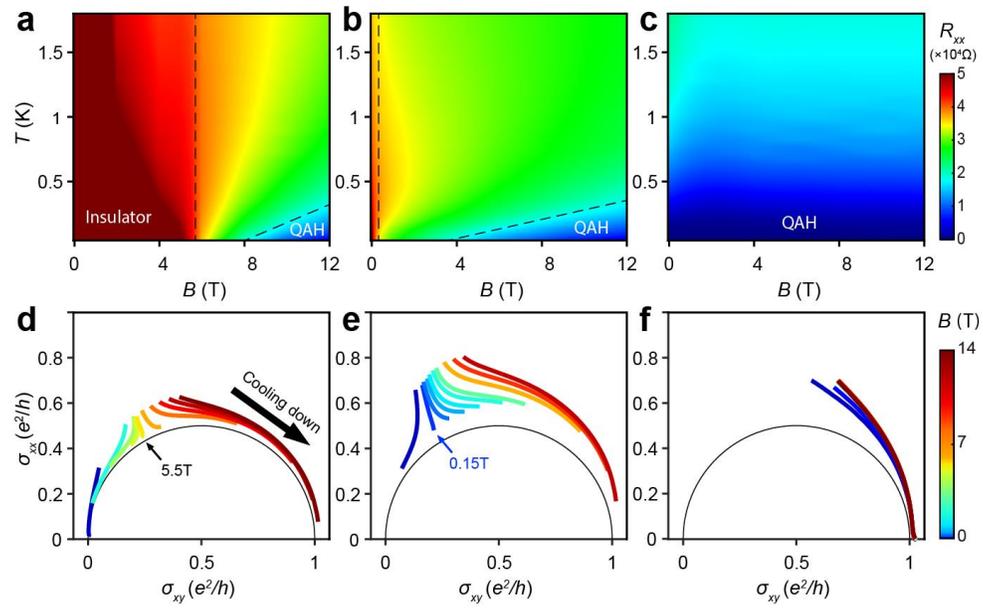

**Figure 3**

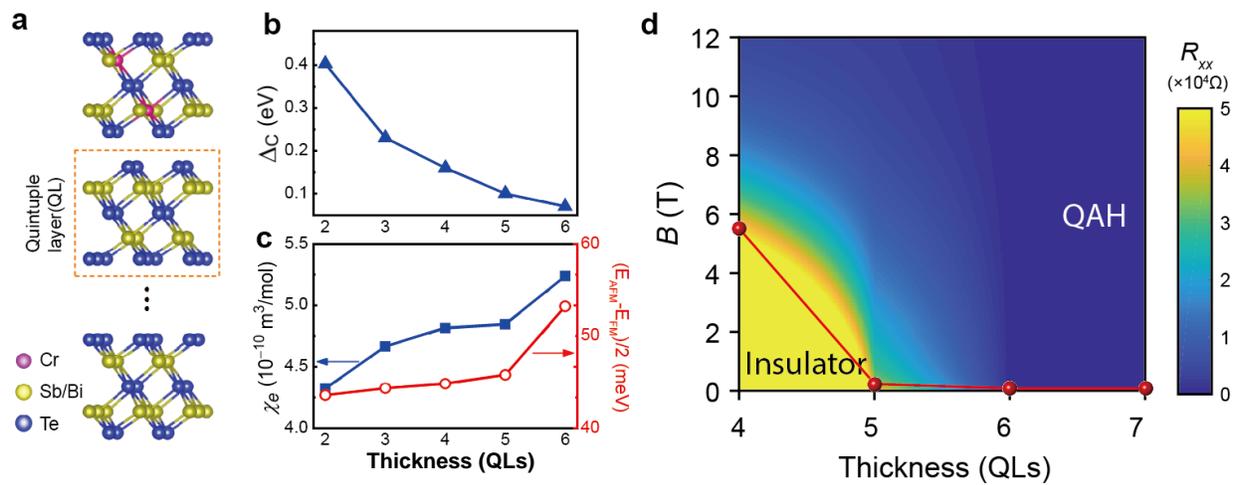

**Figure 4**